\documentclass[aps,preprintnumbers,superscriptaddress,showpacs]{revtex4}
\usepackage{epsfig}
\usepackage{psfrag}
\usepackage{amsfonts}
\usepackage{graphicx}
\usepackage{dcolumn}
\usepackage{bm}

\begin{document}

\title{Entropic destruction of heavy quarkonium from a deformed $AdS_5$ model}

\author{Zi-qiang Zhang}
\email{zhangzq@cug.edu.cn} \affiliation{School of mathematics and
physics, China University of Geosciences(Wuhan), Wuhan 430074,
China}

\author{Zhong-jie Luo}
\email{luozhj@cug.edu.cn} \affiliation{School of mathematics and
physics, China University of Geosciences(Wuhan), Wuhan 430074,
China}

\author{De-fu Hou}
\email{houdf@mail.ccnu.edu.cn} \affiliation{Key Laboratory of
Quark and Lepton Physics (MOE), Central China Normal University,
Wuhan 430079, China}

\author{Gang Chen}
\email{chengang1@cug.edu.cn} \affiliation{School of mathematics
and physics, China University of Geosciences(Wuhan), Wuhan 430074,
China}

\begin{abstract}
In this paper, we study the destruction of heavy quarkonium due to
the entropic force in a deformed $AdS_5$ model. The effects of the
deformation parameter on the inter-distance and the entropic force
are investigated. The influence of the deformation parameter on
the quarkonium dissociation is analyzed. It is shown that the
inter-distance increases in the presence of the deformation
parameter. In addition, the deformation parameter has the effect
of decreasing the entropic force. This results imply that the
quarkonium dissociates harder in a deformed AdS background than
that in an usual AdS background, in agreement with earlier
findings.
\end{abstract}
\pacs{11.25.Tq, 11.15.Tk, 11.25-w}

\maketitle
\section{Introduction}
It is well known that the dissociation of heavy quarknioum can be
regarded as an important experimental signal of the formation of
strongly coupled quark-gloun plasma (QGP) \cite{TMA}. It was
argued earlier that the quarkonium suppression is due to the Debye
screening effects induced by the high density of color charges in
QGP. But the recent experimental research showed a puzzle: the
charmonium suppression at RHIC (lower energy density) is stronger
than that at LHC (larger energy density) \cite{AAD,BBA}.
Obviously, this is in contradiction with the Debye screening
scenario \cite{TMA} as well as the thermal activation through the
impact of gluons \cite{DKH,EV}. To explain this puzzle, some
authors suggested that the recombination of the produced charm
quarks into charmonia may be a solution. This argument was based
on the results \cite{PBR,RLT} that if a region of deconfined
quarks and gluons is formed, the quarkoniums (or bound states) can
be formed from a quark and an antiquark which were originally
produced in separate incoherent interactions. Recently, Kharzeev
\cite{DEK} argued that this puzzle may be related to the nature of
deconfinement and the entropic force would be responsible for
melting the quarknium. This argument originated from the Lattice
results that a large amount of entropy associated with the heavy
quarknioum placed in QGP \cite{DKA1,DKA2,PPE}.

AdS/CFT, which maps a $d$-dimensional quantum field theory to its
dual gravitational theory, living in $d+1$-dimensional, has
yielded many important insights into the dynamics of strongly
coupled gauge theories
\cite{Maldacena:1997re,Gubser:1998bc,MadalcenaReview}. In this
approach, K. Hashimoto et al have studied the entropic destruction
of static heavy quarkonium in $\mathcal{N}=4$ SYM theory and a
confining YM theory firstly. They found that in both cases the
entropy grows as a function of the inter-quark distance giving
rise to the entropic force \cite{KHA}. Soon this studies have been
extended to the case of moving quarkonium \cite{KBF}. It was shown
that the velocity has the effect of increasing the entropic force
thus making the quarkonium melts easier. Recently, we have
analyzed the effect of chemical potential on the entropic force
and observed that the moving quarkonium dissociates easier at
finite density \cite{ZQ}.

Now we would like to give such analyses from AdS/QCD. The
motivation is that AdS/QCD models can provide a nice
phenomenological description of hadronic properties as well as
quark anti-quark interaction, see
\cite{TSS,JEE,AKE,OA3,SH1,UGE,HJ,SH2,OD} and references therein.
In this paper, we will study the entropic force in the
Andreev-Zakharov model \cite{OA3}, one of "soft wall" models. The
Andreev-Zakharov model has some properties: (1) The positive
quadratic term modification in the deformed warp factor
$h(z)=e^{\frac{1}{2}cz^2}$ produces linear behavior of heavy
flavor potential, namely, it can provide confinement at low
temperature. (2) the value of $c$ can be fixed from the $\rho$
meson trajectory, so that the metric contains no free parameter.
Actually, this model has been used to investigate some quantities,
such as thermal phase transition \cite{OA2}, thermal width
\cite{NRF,JSA}, and heavy quark potential \cite{ZQ1}. Likewise, it
is of interest to study the entropic force in this model. Besides
that, we have several other reasons: First, we want to know what
will happen if we have meson in a deformed AdS background? or how
the deformation parameter affects the quarkonium dissociation?
Moreover, evaluation of the entropic force helps us to understand
the "usual" or "unusual" behavior of meson, because one can
compare the results of $c\neq0$ with $c=0$ while the "usual"
behavior of meson can be recovered in the limit $c\rightarrow0$.
On the other hand, such an investigation can be regarded as a good
test of AdS/QCD.

The paper is organized as follows. In the next section, we briefly
review the action of holographic models and then introduce the
Andreev-Zakharov model. In section 3, we study the effects of the
deformation parameter on the inter-distance as well as the
entropic force and then analyze how the deformation parameter
affects the quarkonium dissociation. The last part is devoted to
discussion and conclusion.


\section{the Andreev-Zakharov model}
Before reviewing the Andreev-Zakharov model, let us briefly
introduce the holographic models in terms of the action \cite{OD}
\begin{equation}
S=\frac{1}{16\pi G_5}\int
d^5x\sqrt{g}(\mathcal{R}-\frac{1}{2}(\partial
\phi)^2-V(\phi)-\frac{f(\phi)}{4}F_{MN}F^{MN}),\label{action}
\end{equation}
where $G_5$ is the five-dimensional Newton constant. $g$ denotes
the determinant of the metric $g_{MN}$. $\mathcal{R}$ refers to
the Ricci scalar. $\phi$ is called the scalar and induces the
deformation away from conformality. $f(\phi)$ represents the gauge
kinetic function. $F_{MN}$ stands for the field strength
associated with an Abelian gauge connection $A_M$. $V(\phi)$ is
the potential which contains the cosmological constant term
2$\Lambda$ and some other terms.

To obtain an AdS-black hole space-time, one considers a constant
scalar field $\phi$ (or called dilaton) and assumes that
$V(\phi)=2\Lambda$ as well as $f(\phi)=1$. Then the action of
(\ref{action}) can be simplified as
\begin{equation}
S=\frac{1}{16\pi G_5}\int
d^5x\sqrt{g}(\mathcal{R}-2\Lambda-\frac{1}{4}F_{MN}F^{MN}),\label{action1}
\end{equation}
with the equations of motion
\begin{equation}
\mathcal{R}_{MN}-\frac{1}{2}\mathcal{R}g_{MN}+\Lambda
g_{MN}=\frac{1}{2}(F_{MA}F_N^A-\frac{1}{4}g_{MN}F_{AB}F^{AB})\label{motion},
\end{equation}
\begin{equation}
\nabla_MF^{MN}=0.
\end{equation}
where $\nabla_M$ is the Levi-Civita covariant derivative with
respect to the metric $g_{MN}$.

Supposing that the horizon function $f(z)$ vanishes at the point
$z=z_h$, then the solution of (\ref{motion}) (with vanishing
right-hand side) becomes the $AdS_5$-Schwarzschild metric
\begin{equation}
ds^2=\frac{R^2}{z^2}(-f(z)dt^2+d\vec{x}^2+\frac{1}{f(z)}dz^2),\label{metric1}
\end{equation}
with
\begin{equation}
f(z)=1-\frac{z^4}{z_h^4},
\end{equation}
where $z_h$ can be related to the temperature as $T=1/(\pi z_h)$.
Notice that in the limit $z_h\rightarrow\infty$ (correspond to
zero temperature), the metric of (\ref{metric1}) reduces to the
$AdS_5$ metric, as expected.

To emulate confinement in the boundary theory, one can introduce a
quadratic dilaton, $\phi\propto z^2$, similarly to the
manipulation mentioned in \cite{AKE}. To this end, the
Andreev-Zakharov model can be defined by the metric of
(\ref{metric1}) multiplied by a warp factor,
$h(z)=e^{\phi}=e^{\frac{1}{2}cz^2}$, where $c$ is the deformation
parameter whose value can be fixed from the $\rho$ meson
trajectory as $c\sim0.9GeV^2$ \cite{OA2}. Then the metric of the
Andreev-Zakharov model is given by \cite{OA3}
\begin{equation}
ds^2=\frac{R^2h(z)}{z^2}(-f(z)dt^2+d\vec{x}^2+\frac{1}{f(z)}dz^2).\label{metric2}
\end{equation}

If one works with $r=R^2/z$ as the radial coordinate, the metric
of (\ref{metric2}) turns into
\begin{equation}
ds^2=\frac{r^2h(r)}{R^2}(-f(r)dt^2+d\vec{x}^2)+\frac{R^2h(r)}{r^2f(r)}dr^2,\label{metric}
\end{equation}
with
\begin{equation}
f(r)=1-\frac{r_h^4}{r^4},
\end{equation}
now the wrap factor becomes $h(r)=e^{\frac{cR^4}{2r^2}}$ and the
temperature is $T=r_h/(\pi R^2)$ with $r=r_h$ the horizon. Note
that the two metrics (\ref{metric2}) and (\ref{metric}) are equal
but only with different coordinate systems.

\section{The entropic force}

The entropic force is an emergent force. According to the second
law of thermodynamics, it stems from multiple interactions which
drive the system toward the state with a larger entropy. This
force was originally introduced in \cite{KH} many years ago and
proposed to responsible for the gravity \cite{EP} recently. In a
more recent work, D. E. Kharzeev \cite{DEK} argued that it would
be responsible for dissociating the quarkonium.

In \cite{DEK}, the entropic force is expressed as
\begin{equation}
\mathcal{F}=T\frac{\partial S}{\partial L},\label{f}
\end{equation}
where $T$ is the temperature of the plasma, $L$ represents the
inter-distance of $Q\bar{Q}$, $S$ stands for the entropy.

On the other hand, the entropy is given by
\begin{equation}
S=-\frac{\partial F}{\partial T},\label{s}
\end{equation}
where $F$ is the free energy of $Q\bar{Q}$, which is equal to the
on-shell action of the fundamental string in the dual geometry
from the holographic point of view. In fact, the free energy has
been studied for example in \cite{JMM,ABR,SJR}.

We now follow the calculations of \cite{KHA} to analyze the
entropic force with the metric (\ref{metric}). The Nambu-Goto
action is
\begin{equation}
S=T_F\int d\tau d\sigma\mathcal L=T_F\int d\tau
d\sigma\sqrt{g}\label{S},
\end{equation}
where $T_F=\frac{1}{2\pi\alpha^\prime}$ is the fundamental string
tension and $\alpha^\prime$ can be related to the 't Hooft
coupling constant by $\alpha^\prime=\frac{R^2}{\sqrt{\lambda}}$.
$g$ denotes the determinant of the induced metric with
\begin{equation}
g_{\alpha\beta}=g_{\mu\nu}\frac{\partial
X^\mu}{\partial\sigma^\alpha} \frac{\partial
X^\nu}{\partial\sigma^\beta},
\end{equation}
where $X^\mu$ is the the target space coordinates and $g_{\mu\nu}$
is the metric.

Parameterizing the static string coordinates by
\begin{equation}
X^\mu=(\tau,\sigma,0,0,r(\sigma)),
\end{equation}
one finds the induced metric as
\begin{equation}
g_{00}=\frac{r^2}{R^2}h(r)f(r),\qquad g_{01}=g_{10}=0, \qquad
g_{11}=\frac{r^2}{R^2}h(r)+\frac{R^2}{r^2}h(r)f(r)^{-1}\dot{r}^2,
\end{equation}
with $\dot{r}=\frac{\partial r}{\partial\sigma}$.

Then the Lagrangian density is found to be
\begin{equation}
\mathcal L=\sqrt{a(r)+b(r)\dot{r}^2}\label{L}.
\end{equation}
with
\begin{eqnarray}
&a(r)&=\frac{h^2(r)f(r)r^4}{R^4},\nonumber\\&b(r)&=h^2(r).
\end{eqnarray}

Note that $\mathcal{L}$ does not depend on $\sigma$ explicitly, so
the Hamiltonian density is a constant
\begin{equation}
\mathcal{H}=\mathcal L-\frac{\partial\mathcal
L}{\partial\dot{r}}\dot{r}=constant\label{H}.
\end{equation}

Applying the boundary condition at $\sigma=0$,
\begin{equation}
\dot{r}=0, \qquad r=r_c,
\end{equation}
one finds
\begin{equation}
\mathcal{H}=\sqrt{\frac{h^2(r_c)f(r_c)r_c^4}{R^4}}\label{H1},
\end{equation}
with
\begin{equation}
f(r_c)=1-\frac{r_h^4}{r_c^4}, \qquad
h(r_c)=e^{\frac{cR^4}{2r_c^2}},
\end{equation}
where $r_c$ is the lowest position of the string in the bulk.

From (\ref{L}), (\ref{H}) and (\ref{H1}), one gets
\begin{equation}
\dot{r}=\frac{dr}{d\sigma}=\sqrt{\frac{a^2(r)-a(r)a(r_c)}{a(r_c)b(r)}}\label{dotr}.
\end{equation}

By integrating (\ref{dotr}), the inter-quark distance is obtained
\begin{equation}
L=2\int_{r_c}^{\infty}dr\sqrt{\frac{a(r_c)b(r)}{a^2(r)-a(r)a(r_c)}}\label{x},
\end{equation}
with
\begin{equation}
a(r_c)=\frac{h^2(r_c)f(r_c)r_c^4}{R^4}.
\end{equation}

\begin{figure}
\centering
\includegraphics[width=8cm]{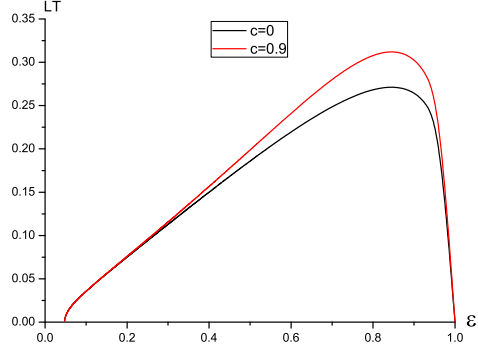}
\caption{$LT$ versus $\varepsilon$ with $\varepsilon\equiv
r_h/r_c$. From top to bottom $c=0.9GeV^2$ and $c=0$ respectively.
Here we take $R=1$.}
\end{figure}

To analyze the effect of the deformation parameter on the
inter-distance, we plot $LT$ as a function of $\varepsilon$ with
$\varepsilon\equiv r_h/r_c$ for $c=0$ and $c=0.9GeV^2$ in Fig.1.
From the figures, one can see that $LT$ increases in the presence
of $c$. Namely, the deformation parameter has the effect of
increasing the inter-distance.

Moreover, one finds that for each plot $LT$ is a increasing
function for $\varepsilon<\varepsilon_{max}$ but a decreasing one
for $\varepsilon>\varepsilon_{max}$. In fact, in the later case
some new configurations \cite{DB} should be taken into account.
However, these configurations are not solutions of the Nambu-Goto
action so that the range of $LT>LT_{max}$ is not trusted. In other
words, we have more interest in the range of $LT<LT_{max}$. For
convenience, we write $b=LT_{max}$. With numerical methods, we
find $b\simeq0.31$ for $c=0.9GeV^2$ and $b\simeq0.27$ for $c=0$.

Next we discuss the free energy. There are two cases:

1). If $L>\frac{b}{T}$, the fundamental string will break in two
pieces implying the quarks are completely screened. For this case,
the choice of the free energy $F^{(2)}$ is not unique \cite{MCH},
we here choose a configuration of two disconnected trailing drag
strings \cite{CPH}, that is
\begin{equation}
F^{(2)}=2T_F\int_{r_h}^{\infty} h(r)dr.
\end{equation}

In terms of (\ref{s}), one finds
\begin{equation}
S^{(2)}\simeq e^{\frac{c}{2r_h^2}}(1-\frac{c}{
r_h^2})\sqrt{\lambda}\theta(L-\frac{b}{T})\label{S2},
\end{equation}
notice that the results of \cite{KHA} can be reproduced if one
neglects the effect of the reformation parameter by plugging $c=0$
in (\ref{S2}).

2). If $L<\frac{b}{T}$, the fundamental string is connected. The
free energy of the quark anti-quark pair can be obtained by
substituting (\ref{dotr}) into (\ref{S}), that is
\begin{equation}
F^{(1)}=\frac{1}{\pi\alpha^\prime}\int_{r_c}^{\infty} dr
\sqrt{\frac{a(r)b(r)}{a(r)-a(r_c)}}.
\end{equation}

Likewise, using (\ref{s}) one finds
\begin{equation}
S^{(1)}=-\frac{\sqrt{\lambda}}{2\pi}\int_{r_c}^{\infty}
dr\frac{[a^\prime(r)b(r)+a(r)b^\prime(r)][a(r)-a(r_c)]-a(r)b(r)[a^\prime(r)-a^\prime(r_c)]}{\sqrt{a(r)b(r)[a(r)-a(r_c)]^3}},\label{S1}
\end{equation}
where the derivatives are with respect to $r_h$ and we have used
the relation $\alpha^\prime=\frac{R^2}{\sqrt{\lambda}}$.

To analyze the effect of the deformation parameter on the entropic
force, we plot $S^{(1)}/\sqrt{\lambda}$ as a function of $LT$ for
$c=0$ and $c=0.9GeV^2$ in Fig 2, respectively. One can see that
increasing $c$ leads to smaller entropy at small distances. In
addition, from (\ref{f}) one knows that the entropic force is
related to the growth of the entropy with the distance, so one
finds that increasing $c$ leads to decreasing the entropic force.
On the other hand, the entropic force is responsible for melting
the quarkonium. Thus, one concludes that the presence of the
deformation parameter tends to decrease the entropic force thus
making the heavy quarkonium dissociates harder. This results can
be understood as follows. Increase of the inter-distance can be
regarded as decrease of $r_h$ or decrease of the system
temperature. Since the deformation parameter has the effect of
increasing the inter-distance, it will cool the system temperature
thus making the quarkonium dissociates harder. Interestingly, it
was argued \cite{JSA} that the deformation parameter has the
effect of increasing the thermal thus increasing the dissociation
length, in agreement with our findings.

\begin{figure}
\centering
\includegraphics[width=8cm]{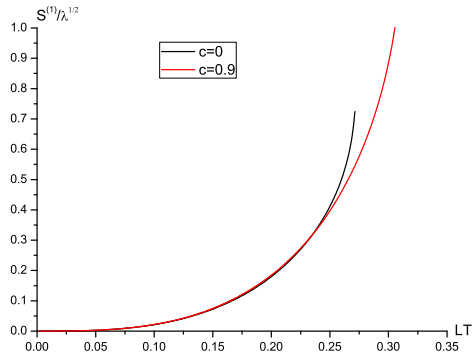}
\caption{$S^{(1)}/\sqrt{\lambda}$ against $LT$. Here we take
$R=1$.}
\end{figure}

\section{summary and discussions}

In heavy ion collisions, the dissociation of heavy quarknioum is
an important experimental signal for QGP formation. Recently, the
destruction of heavy quarkonium due to the entropic force has been
discussed in the context of AdS/CFT \cite{KHA}. It was shown that
a sharp peak of the entropy exists near the deconfinement
transition and the growth of the entropy with the distance is
responsible for the entropic force.

In this paper, we have investigated the destruction of heavy
quarkonium in a deformed $AdS_5$ model. The effect of the
deformation parameter on the inter-distance was analyzed. The
influence of the deformation parameter on the entropic force was
also studied. It is shown that the inter-quark distance increases
in the presence of the deformation parameter. Moreover, the
deformation parameter has the effect of decreasing the entropic
force. Since the entropic force is responsible for destroying the
bound quarkonium states, therefore, we conclude that the presence
of the deformation parameter tends to decrease the entropic force
thus making the quarkonium melts harder, consistently with the
findings of \cite{JSA}. Also, we have presented a possible
understanding to this result: Increase of the inter-distance is
equivalent to decrease of $r_h$ or decrease of the system
temperature. As the deformation parameter can increase the
inter-distance, it will cool the system temperature thus making
the quarkonium dissociates harder.

In addition, to understand the "usual" or "unusual" behavior of
meson, we have compared the results between $c\neq0$ and $c=0$. It
is found that the quarkonium dissociates harder in a deformed AdS
background than that in an usual AdS background.

Finally, it would be interesting to study the entropic force in
some other holographic QCD models, such as the Sakai-Sugimoto
model \cite{TSS} and the Pirner-Galow model \cite{HJ}. This will
be left as a further investigation.

\section{Acknowledgments}

This research is partly supported by the Ministry of Science and
Technology of China (MSTC) under the ¡°973¡± Project no.
2015CB856904(4). Zi-qiang Zhang and Gang Chen are supported by the
NSFC under Grant no. 11475149. De-fu Hou is supported by the NSFC
under Grant no. 11375070, 11521064.


\end{document}